\documentclass[twocolumn,showpacs,preprintnumbers,amsmath,amssymb,prb,aps,nofootinbib]{revtex4-1}
\usepackage{graphicx}
\usepackage{dcolumn}
\usepackage{bm}
\begin{document}


\title{Raman studies on amorphous carbon layers $-$ Raman-Untersuchungen von amorphen Kohlenstoffschichten}

\author{J. Debus} \email{joerg.debus@tu-dortmund.de}
\affiliation{Experimentelle Physik 2, Technische Universit\"at Dortmund, 44227 Dortmund, Germany}

\begin{abstract}
Raman spectroscopic study of amorphous carbon layers for two different excitation wavelengths at room temperature. The amount of sp$^3$ bondings is estimated to about 10\% for both samples, evaluated from the ratio of the D- and G-Raman line intensities. The properties of the bondings in the two samples are discussed.

Raman-spektroskopische Untersuchungen von amorphen Kohlenstoffschichten f\"ur zwei unterschiedliche Anregungswellenl\"angen bei Raumtemperatur. Der sp$^3$-Bindungsanteil is zu 10\% f\"ur beide Proben bestimmt worden, ermittelt aus dem Intensit\"atsverh\"altnis der D- and G-Raman-Linien. Die Eigenschaften der Bindungen in den beiden untersuchten Proben werden diskutiert.
\end{abstract}

\maketitle

\subsection{Theoretisches: Amorpher Kohlenstoff}
Die Raman-Spektroskopie kann zur Unterscheidung der sp$^2$- und sp$^3$-Hybridorbitale in diamant\"ahnlichem Kohlenstoff dienen, da ihre entsprechenden Raman-Signale in Diamand und Graphit voneinander hinrei\-chend separiert sind ($1350$ und $1580\,$cm$^{-1}$). Die Raman-Spektroskopie im sichtbaren Spektralbereich ist nur von eingeschr\"anktem Nutzen, da in dem sichtbaren Spektralbereich der Streuquerschnitt der sp$^2$-Orbitale am gr\"o\ss ten ist und der Beitrag der sp$^3$-Orbitale in den Raman-Spektren \"uberdeckt wird. Die aufgenommenen Spektren spiegeln die Konfiguration/ Ordnung der sp$^2$-Gitterpl\"atze wider, \"uber die sp$^3$-Gitterpl\"atze kann nur indirekt eine Aussage getroffen werden. Ihr Anteil in der Kohlenstoffprobe l\"asst sich mittels des von Ferrari und Robertson vorgeschlagenen 3-Stufen-Modells, das sich auf das dispersive Verhalten des G- und D-Peaks und ihr Intensit\"atsverh\"altnis bezieht, herleiten\cite{ferrari}. Graphit hat eine einzelne aktive Raman-Mode bei $1580\,$cm$^{-1}$, sie entspricht der Mode in der Mitte der Brillouin-Zone mit E$_{2g}$-Symmetrie. Sie wird mit ''G'' (f\"ur Graphit) bezei\-chnet. Graphit mit ungeordneter Struktur besitzt eine zweite Mode bei etwa $1350\,$cm$^{-1}$ mit A$_{1g}$-Symmetrie, ihre Kennzeichnung ist ''D'' (f\"ur \textit{disorder}). Die G-Mode von Graphit ist eine oszillierende ebene Bewegung (Dehnung der Hybridbindung) eines jeden Kohlenstoffpaares, dessen Atome sp$^2$-Bindungen eingehen. Sie k\"onnen sowohl in Ketten als auch in konjugiert-planaren Ringen auftreten. Die D-Mode hingegen ist durch eine ''atmende'' Mode von sp$^2$-Gitterpl\"atzen, angeordnet in sechsf\"altigen Ringen (nicht in Ketten), bedingt. 

Kohlenstoff tritt in unterschiedlichen Konfigurationen auf, kategorisierbar in kristalline und ungeordnete Strukturen, da die Kohlenstoff-Orbitale unterschiedlich hybridisieren k\"onnen: Es sind sp$^3$-, sp$^2$- und sp$^1$-Hybridorbitale m\"oglich. In der sp$^3$-Konfiguration (wie in Diamant) ist jedes der vier Valenzelektronen eines Kohlenstoffatoms \"uber sp$^3$-Orbitale tetraedrisch koordiniert. Dies bedingt eine starke $\sigma$-Bindung zu einem benachbarten Atom. In der dreifach-koordinierten sp$^2$-Konfiguration (bspw. in Graphit) nehmen drei der vier Valenzelektronen trigonal-gerichtete sp$^2$-Orbitale ein, welche $\sigma$-Bindungen in einer Ebene bilden. Das vierte Elektron eines sp$^2$-Atoms liegt in einem $\pi$-Orbital\cite{robertson}, das orthogonal zu der $\sigma$-Bindungsebene liegt\cite{remark}.

Die Raman-Spektren von amorphen Materialien sollten die gesamte phononische Zustandsdichte widerspiegeln im Hinblick auf die Relaxierung der Auswahlregeln des Phonon-\textbf{k}-Vektors. In sichtbaren Raman-Messungen entspricht die Anregungsenergie von $2.3\,$eV ($532\,$nm) dem $\pi$-$\pi^*$-\"Ubergang der sp$^2$-Hybridorbitale. Dies f\"uhrt zu einer resonanten Erh\"ohung des Raman-Streuquerschnitts. Dieses Problem kann umgangen werden, indem eine Anregung im ultravioletten Bereich erfolgt. Hierbei sollten die $\sigma$-Zust\"ande der sp$^2$- und sp$^3$-Orbitale angeregt werden, eine entsprechend hohe Anregungsenergie ist zu w\"ahlen ($\approx 5\,$eV, $260\,$nm). Sofern die Gitterperiodizit\"at verloren geht und die \textbf{k}-Vektor-Auswahlregel von optischen und phononischen \"Uberg\"angen nicht mehr g\"ultig ist, entspricht ein Raman-Spektrum eines amorphen Netzwerks der phononischen Zustandsdichte, gewichtet mit einem passend zu w\"ahlenden Matrixelement. Beschrieben wird dieser Zusammenhang in der Shuker-Gammon-Gleichung\cite{ferrari}. Ein Grund f\"ur die Dominanz der G- und D-Moden in den Raman-Spektren von amorphem Kohlenstoff ist die \"uberwiegend auftretende Streuung in den sp$^2$-Hybridorbitalen. Die $\pi$-Zust\"ande liegen bei geringerer Energie als die $\sigma$-Niveaus; sie sind auch wesentlich st\"arker polarisierbar. Hierdurch ist der sp$^2$-Streuquerschnitt zwei bis drei Gr\"o\ss enordnungen gr\"o\ss er als der der sp$^3$-Orbitale. Dennoch folgen die Raman-Spektren nicht nur der Phononzustandsdichte der sp$^2$-Gitterpl\"atze. Der Einfluss des oben genann\-ten Matrixelements ist in $\pi$-gebundenen Netzwerken wahrscheinlicher als bei $\sigma$-Bindungen. Damit werden die Raman-Spektren durch den Grad der Ordnung der sp$^2$-Gitterpl\"atze und nicht durch den sp$^2$-Anteil gepr\"agt.

\subsection{Experimentelle Details}
Die Raman-Messungen wurden mit einem gepulsten Nd:YVO-Laser, betrieben in der zweiten oder dritten Harmonischen, durchgef\"uhrt. Die Raman-Signale wurden mit einem dreistufigen Monochromator und einer stickstoffgek\"uhlten CCD-Kamera aufgenommen. Der Tripel-Monochromator arbeitete in dem subtraktiven Modus, in dem die ersten beiden Stufen als Laserfilter dienten und die Dispersion ausschlie\ss lich von der dritten/letzten Stufe \"ubernommen wurde. Die maximale spektrale Aufl\"osung dieser Konfiguration lag unterhalb von $1\,$cm$^{-1}$. Der st\"orende Einfluss von Laserstreulicht konnte zus\"atzlich mittels gekreuzter Polarisatoren verringert werden. Die Messungen erfolg\-ten bei Raumtemperatur.

\subsection{Ergebnisse}
Die Charakterisierung von diamant\"ahnlichem Kohlenstoff durch sichtbare Raman-Messungen basiert auf empirischen Regeln (3-Stufen-Modell), die sich auf das dispersive Verhalten des Diamant- und Graphit-Signals und ihr Intensit\"atsverh\"altnis beziehen\cite{ferrari, robertson}. Eine ausf\"uhrliche Beschreibung kann auch in einer Ver\"offentlichung von Ferrari und Robertson aus dem Jahr 2001 gefunden werden\cite{ferrarb}. Im Folgenden wird das 3-Stufen-Modell auf die gemessenen Beobachtungen angewandt.

\begin{figure}[t]
\centering
\includegraphics[width=6.8cm]{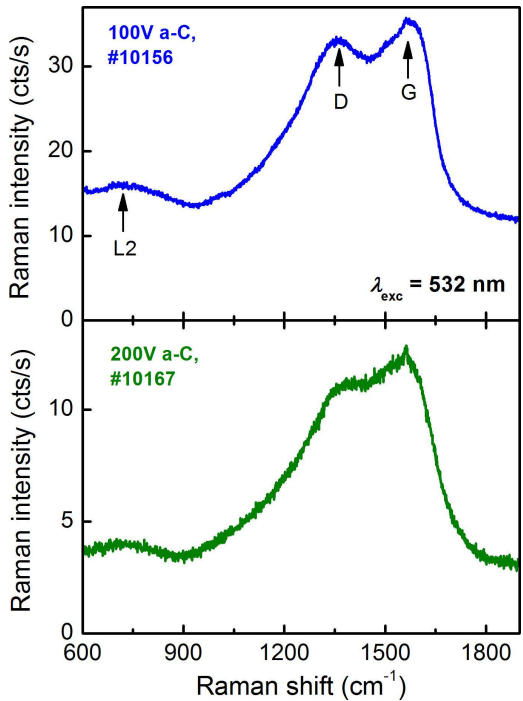}
\caption{Raman-Spektren der a-C-Proben \#10156 und \#10167 f\"ur die Anregung mit Licht der Wellenl\"ange $532\,$nm und mit einer mittleren Leistungsdichte von $30\,$W/cm$^2$. Neben dem G- und D-Peak deutet sich bei etwa $750\,$cm$^{-1}$ ein weiteres Signal an. Dieses wird in der Literatur mit L2 bezeichnet.}
\label{fig:fig1}
\end{figure}

\begin{figure}[t]
\centering
\includegraphics[width=6.8cm]{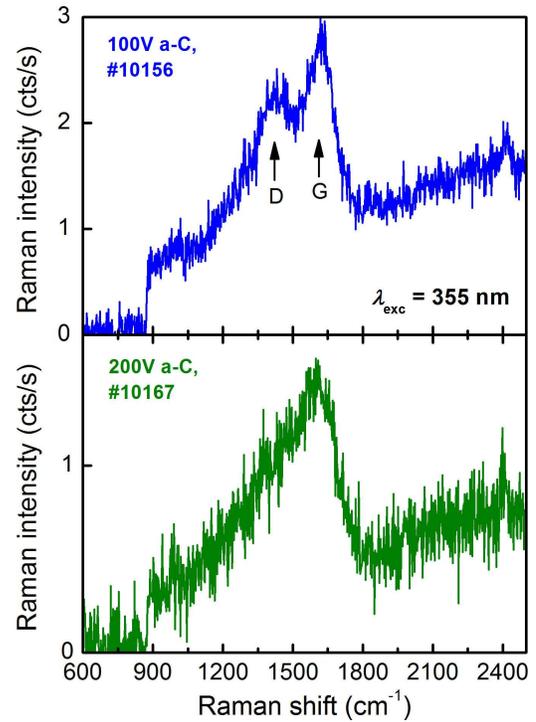}
\caption{Raman-Spektren der a-C-Proben \#10156 und \#10167 f\"ur $355\,$nm Anregung, die mittlere Leistungsdichte betr\"agt $11\,$W/cm$^2$. Im Vergleich zu der Anregung mit $532\,$nm ist die Intensit\"at der Raman-Signale um den Faktor vier geringer. Bei $2400\,$cm$^{-1}$ scheint sich ein weiterer Peak anzudeuten.}
\label{fig:fig2}
\end{figure}

In Abbildung \ref{fig:fig1} sind die Raman-Spektren f\"ur die amorphen Kohlenstoffproben (a-C) \#10156 und \#10167 bei einer Anregung von $532\,$nm und einer mittleren Leistungsdichte von etwa $30\,$W/cm$^2$ gezeigt. Die Intensit\"aten der Raman-Signale sind auf $1\,$sek.-Detektionszeit normalisiert. Aus den Spektren gehen sowohl die beiden typischen D- und G-Signale hervor, als auch ein wei\-terer Peak bei rund $750\,$cm$^{-1}$. Letzterer, charakteristisch f\"ur a-C-Proben, l\"asst sich auf einen geringen Anteil an sp$^3$-Bindungen zur\"uckf\"uhren\cite{ferrarb}. In der 100V-a-C-Probe \#10156 scheinen der D- und G-Peak st\"arker voneinander separiert zu sein. F\"ur eine UV-Anregung mit $355\,$nm (s. Abb. \ref{fig:fig2}) nimmt die Intensit\"at der Raman-Signale ab, unter Beachtung der unterschiedlichen Laserleistungen. Au\ss erdem scheint sich um $2400\,$cm$^{-1}$ ein weiteres Signal abzuzeichnen, dies ist ein Anzeichen f\"ur die Amorphisierung der Kohlenstoffbindungen\cite{ferrari}. Die eindeutigere spektrale Separation des D- und G-Peaks ist in der 100V-a-C-Probe weiterhin gegeben.

\begin{figure}[t]
\centering
\includegraphics[width=6.8cm]{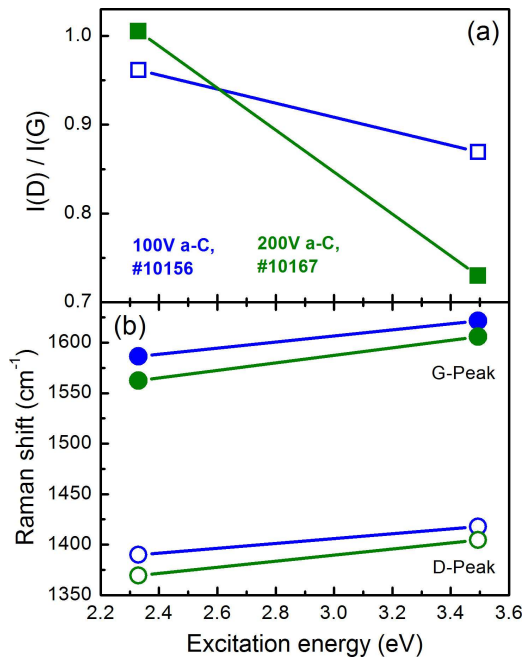}
\caption{Intensit\"atsverh\"altnis des D- und G-Peaks sowie deren Raman-Verschiebungen f\"ur die amorphen Kohlenstoffproben \#10156 und \#10167. Das Verh\"altnis $I(D)/I(G)$ basiert auf den Amplituden der Raman-Signale, die Peak-Positionen wurden mittels Lorentz-Kurven bestimmt. Merkmale des D-Peaks werden mit offenen Symbolen gekenn\-zeichnet. W\"ahrend die Intensit\"atsverh\"altnisse ein kontr\"ares Verhalten zeigen, deuten die Raman-Verschiebungen einen li\-nearen Verlauf mit nahezu demselben \"Anderungskoeffizienten an. Bei der Anregung im Ultravioletten nimmt die G-Peak-Verschiebung ab, da mit UV-Anregung die Kohlenstoff-Cluster st\"arker angesprochen werden.}
\label{fig:fig3}
\end{figure}

In der Abbildung \ref{fig:fig3}~(a) ist das Intensit\"atsverh\"altnis des D- und G-Peaks, gegeben \"uber die Peak-Amplituden, in Abh\"angigkeit von der Anregungsenergie aufgetragen. Die Raman-Verschiebung der Peaks f\"ur die beiden verwendeten Anregungsenergien ist in Abbildung \ref{fig:fig3}~(b) gezeigt, die Raman-Signale wurden mit Lorentz-Kurven angepasst. Typischerweise wird der D-Peak mit einer Lorentz-Kurve und der G-Peak mit der Breit-Wigner-Fano-Kurve angepasst\cite{ferrari}. W\"ahrend die Intensit\"at des D-Peaks gegen\"uber der des G-Peaks mit steigender Anregungsenergie geringer wird und diese Abnahme f\"ur die 200V-Probe \#10167 st\"arker ausgepr\"agt ist (von $I(D)/I(G)\approx 0.95$ auf $0.73$), vergr\"o\ss ert sich die Raman-Verschiebung gleicherma\ss en f\"ur beide Peaks (ca. um $2,5\%$). Die dispersive \"Anderung der Raman-Verschiebung unterscheidet sich f\"ur den D- und G-Peak, prozentual steigt der G-Peak um etwa $30\%$ st\"arker an. In nahezu gleichem Ma\ss e unterscheiden sich die Raman-Verschiebungen f\"ur die beiden Proben. Anders ausgedr\"uckt, s. Abb. \ref{fig:fig3}~(b), die Differenz $\Delta$ der Verschiebungen f\"ur die h\"ohere Anregungsenergie scheint kleiner auszu\-fallen, als es f\"ur die geringere Anregungsenergie der Fall ist.

Mittels der Diagramme und des Bezugs zu dem 3-Stufen-Modell k\"onnen einige Schlussfolgerungen gezogen werden: Sowohl der D- als auch der G-Peak verhalten sich dispersiv, wobei der Grad der Raman-Verschiebung f\"ur die beiden Proben unterschiedlich ist. Die Raman-Verschiebungen nehmen mit steigender Anregungsenergie zu. Ob der Verlauf linear ist, l\"asst sich aufgrund der geringen Anzahl an Anregungsenergien nicht determinieren. Der Grad der Ver\"anderung der Raman-Verschiebung kann verwendet werden, um den Anteil der sp$^3$-Bindungen zu beschreiben. Aus den Positionen der G-Peaks l\"asst sich ablesen, dass f\"ur die 100V-Probe ungef\"ahr $5\%$ sp$^3$-Bindungen vorliegen, w\"ahrend bei der 200V-Probe der Anteil ungef\"ahr $10\%$ betr\"agt. Das Verh\"altnis $I(D)/I(G)$ l\"asst f\"ur $E_{\text{exc}}=2.3\,$eV auf einen Anteil von etwa $10\%$ schlie\ss en. Nach dem ph\"anomenologischen 3-Stufen-Modell, speziell der Amorphisierungstrajektorie, gilt f\"ur amorphen Kohlenstoff, dass ein hoher sp$^3$-Anteil vorliegt (etwa $20\%$), sofern das Verh\"altnis $I(D)/I(G)$ gegen null geht. Da in dem vorliegenden Fall der Wert um $1$ tendiert (bei $E_{\text{exc}}=2.3\,$eV), liegt der Anteil zwischen dem von nanokristallinem Kohlenstoff ($0\%$) und amorphem Kohlenstoff ($20\%$), d.h. bei etwa $10\%$. In erster N\"aherung gilt f\"ur beide Proben:
\begin{equation}
\text{sp}^3\text{-Anteil} = (10\pm 4)\% \quad . 
\end{equation}
Dennoch sollte ber\"ucksichtigt werden, dass die beiden Proben ein unterschiedliches Verh\"altnis von sp$^2$- und sp$^3$-Bindungen besitzen.

Ein Raman-Spektrum von amorphem Kohlenstoff h\"angt von mehreren spezifischen Eigenschaften der sp$^2$- und sp$^3$-Bindungen ab\cite{ferrari}: Die Cluster-Formation in der sp$^2$-Phase, die Bindungs-(Un)Ordnung, die Existenz von sp$^2$-Ringen und Ketten sowie das sp$^2$/sp$^3$-Verh\"altnis. Diese Faktoren beeinflussen das Aussehen sowie das Verhalten der D- und G-Raman-Signale. Im Rahmen des 3-Stufen-Modells sind die Proben der Stufe 2 (\"Ubergang von nanokristallinem Graphit zu amorphem Kohlenstoff) zu zuschreiben. In dieser Stufe f\"uhren Defekte in den Graphitschichten zu einem Aufweichen der Phonon-Moden, insbesondere des G-Peaks. Die Shuker-Gammon-Gleichung ist g\"ultig und die phononische Gesamtzustandsdichte besitzt keine relevante \"Ahnlichkeit mit der von Graphit. Die Proben sind charakterisiert durch eine nahezu vollst\"andig ungeordnete Struktur, die sich vornehmlich aus sp$^2$-Bindungen zusammensetzt. Eine gr\"o\ss ere Anzahl von sechsf\"altigen Kohlenstoffringen liegt vor, die Anzahl der Kettenbindungen ist gering\cite{robertson}. 

Das dispersive Verhalten des G-Peaks ist charakteristisch f\"ur ungeordneten Kohlenstoff, wobei die Dispersion proportional zu dem Grad der Unordnung ist. In Graphit kann der G-Peak nicht dispergieren, da er die Raman-aktive Phonon-Mode ist. In nanokristallinem Kohlenstoff nimmt die Raman-Verschiebung des G-Peaks aufgrund des Phonon-Confinements geringf\"ugig zu. Aller\-dings zeigt sich bei variierender Anregungsenergie ein nicht-dispersives Verhalten des G-Peaks, da er auch in nanokristallinem Kohlenstoff die phononische Zu\-standsdichte widerspiegelt. Sofern die Unordnung in der Kohlenstoffprobe zunimmt, tritt eine Verschiebung des G-Peaks mit der Anregungsenergie auf. In diesem Fall liegt eine gr\"o\ss ere Anzahl an Konfigurationen mit unterschiedlichen lokalen Bandl\"ucken und unterschiedlichen Phonon-Moden vor. Die Dispersion ist bedingt durch eine resonante Auswahl von sp$^2$-Konfigurationen oder Clustern mit gr\"o\ss eren $\pi$-Bandl\"ucken and damit h\"oheren Schwingungsfrequenzen.

Ein weiterer interessanter Effekt, der Trend-Inversion bei UV-Anregung genannt wird\cite{ferrarb}, deutet sich bei den untersuchten Proben nicht an. Sofern zwei Proben eine \"ahnliche G-Peak-Position in sichtbaren Raman-Spektren besitzen, aber sehr unterschiedliche in UV-Raman-Spektren, dann tritt in der Probe mit der niedrigeren G-Peak-Position im Ultravioletten eine st\"arkere sp$^2$-Cluster-Bildung auf. Mit diesem inversen Verhalten, abh\"angig von der Anregungsenergie, l\"asst sich die St\"arke der Cluster-Formation in der sp$^2$-Phase bestimmen. Des Weiteren ist das Verh\"altnis $I(D)/I(G)$ gr\"o\ss er f\"ur eine h\"ohere Cluster-Konzentration. Dies erkl\"art die Abnahme von $I(D)/I(G)$ bei steigender Anregungsenergie. Sehr ungeordneter amorpher Kohlenstoff hat eine spezifische Gr\"o\ss enverteilung von Clustern und kann nicht alle atmenden Modenfrequenzen der Ring-Cluster beliebiger Gr\"o\ss e umfassen. Daher kann von einem nahezu konstanten Verh\"altnis $I(D)/I(G)$ und einer festen Position des D-Peaks f\"ur absolut amorphen Kohlenstoff ausgegangen werden. Bez\"uglich der Strukturcharakteri\-sierung sind die untersuchten Proben daher zwischen nanokristallinem Graphit und amorphem Kohlenstoff einzuordnen. Gem\"a\ss\ der vorherigen Ausf\"uhrungen scheint die Cluster-Bildung in der 200V-Probe st\"arker ausgepr\"agt zu sein.

\subsection{Fazit und Ausblick}
Der Anteil der sp$^3$-Bindungen in den beiden untersuchten a-C-Proben konnte relativ genau zu $10\%\pm 4\%$ bestimmt werden. Definitive Aussagen \"uber das dispersive Verhalten des G- und D-Peaks erfordern eine gr\"o\ss ere Anzahl von Anregungsenergien; ein linearer Zusammenhang zwischen der Raman-Verschiebungen und der Anregungsenergie muss nicht vorliegen. Die durchgef\"uhrten Raman-Messungen waren aus experimenteller Sicht n\"utzlich und k\"onnen als Basis f\"ur zuk\"unftige Untersuchungen von Kohlenstoffproben die\-nen.

Neben der Anregung mit $532\,$nm und $355\,$nm wurden die Proben auch mit der Wellenl\"ange $668\,$nm (Farbstoff-Laser) und $748\,$nm (Titanium-Saphir-Laser) angeregt. Allerdings \"uberlagerte die Hintergrund-Lumineszenz bzw. das Streulicht der Laser die Raman-Signale der Proben. Die Verwendung von gekreuzten linearen Polarisatoren konnte die Lumineszenz einigerma\ss en unterdr\"ucken, eindeutig ungest\"orte Spektren konn\-ten dennoch nicht erhalten werden. Im Hinblick auf weitere Messungen kann eine bessere Optimierung der Justage des dreistufigen Monochromators dieses Streulicht-Problem unterbinden.

Ein weiteres Problem k\"onnte eine zu geringe Laserleistung darstellen. Die in diesem Zwischenbericht dargelegten Spektren wurden mit einem gepulsten Laser erzeugt, dessen integrale Leistung h\"oher als die eines kontinuierlichen Lasers ist. Hier w\"are der Einsatz eines durchstimmbaren gepulsten Lasers vorteilhaft, um letzt\-lich die Aufnahmezeit der Raman-Spektren zu verringern.

Die Raman-Spektren im sichtbaren Spektralbe\-reich h\"angen haupts\"achlich von der Ordnung der sp$^2$-Gitterpl\"atze ab, eine Aussage \"uber den Anteil der sp$^3$-Pl\"atze ist nur indirekt m\"oglich. Eine konkrete Aussage \"uber die sp$^3$-Bindungen ist ausschlie\ss lich f\"ur hohe energetische Anregung ($\approx 255\,$nm) m\"oglich. Steigt die Anregungsenergie, treten zwei Effekte auf\cite{gilkes}: Die Anregung von solchen sp$^2$-Konfigurationen mit einer breiten energetischen L\"ucke, und, sofern die Anregung im tiefen Ultravioletten liegt, die Anregung von Moden der $\sigma$-Zust\"ande der Kohlenstoffbindungen. F\"ur zuk\"unftige Messungen sollte daher die Anregung im tiefen Ultravioletten in die Betrachtungen miteinbezogen werden.

\end{document}